\documentclass[a4paper,11pt]{article}
\usepackage{pos}

\title{High-energy cosmic particles}

\author{Silvia Mollerach}

\affiliation{
Centro At\'omico Bariloche, Comisi\'on Nacional de Energ\'\i a At\'omica\\ 
Consejo Nacional de Investigaciones Cient\'\i ficas y T\'ecnicas (CONICET)\\ 
Av. Bustillo 9500, R8402AGP, Bariloche, Argentina}

\emailAdd{mollerach@cab.cnea.gov.ar}

\abstract{A review of the status of the knowledge in the field of High-energy cosmic particles is presented.
The spectrum, arrival direction distribution and composition measurements are summarized, to-
gether with some implications for the understanding of the cosmic ray origin and their propagation.
Special emphasis is put in the ultra-high energy range, corresponding to particles of extragalactic
origin.

 }

\FullConference{%
  40th International Conference on High Energy physics - ICHEP2020\\
  July 28 - August 6, 2020\\
  Prague, Czech Republic (virtual meeting)
}

\begin{document}
\maketitle

\section{Introduction}
\begin{figure}[h]
\centering{
\includegraphics[scale=.1,angle=0]{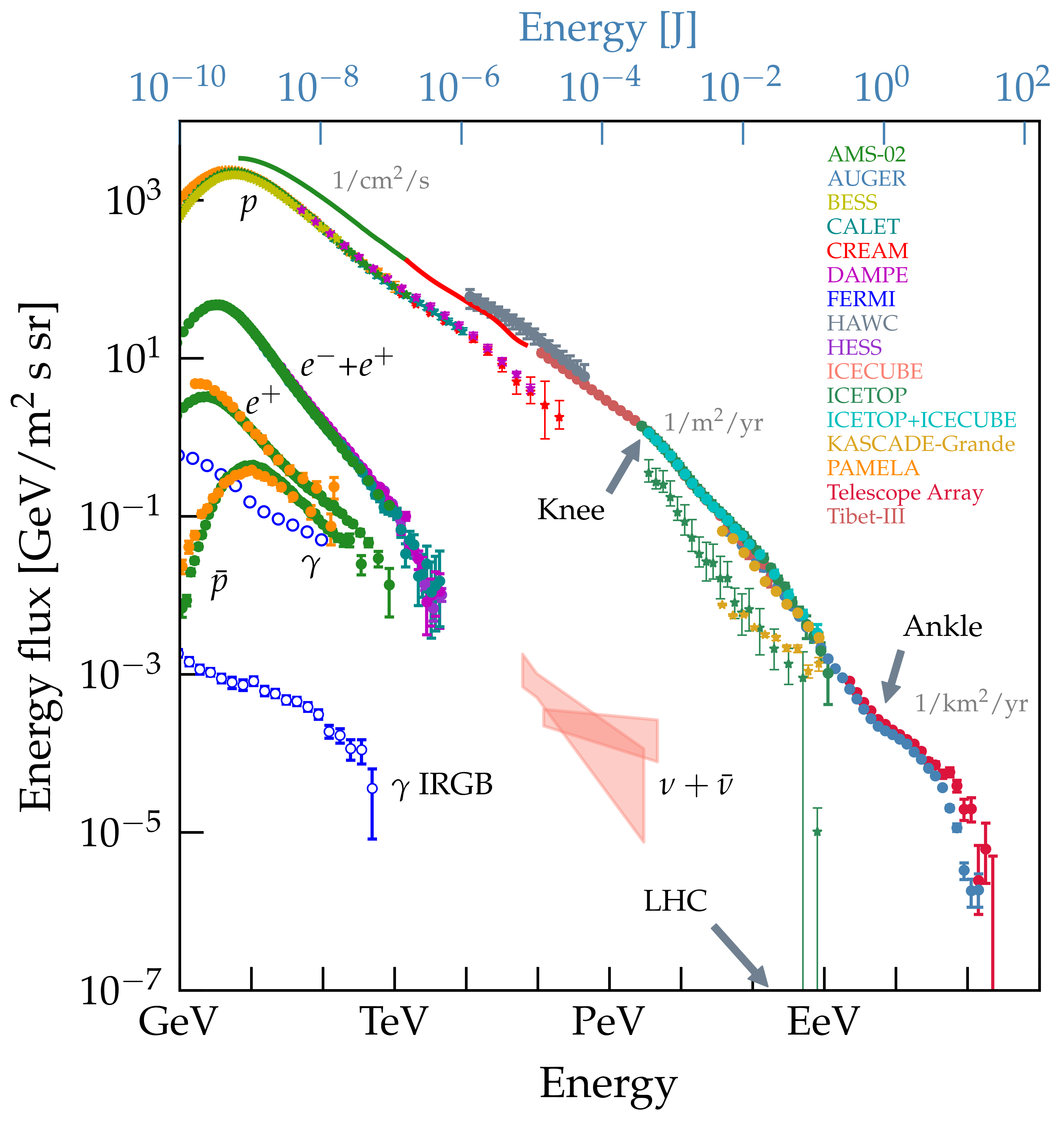}}
\caption{Energy flux ($E^2 d\Phi/dE$) of cosmic particles as measured by different experiments. The total spectrum, as well as those of different components, are included (courtesy of C. Evoli \cite{evoli}).}
\label{fig:allspec}
\end{figure}

High energy cosmic rays (CR) are particles that arrive from space with a wide range of energies from $10^9$ eV to $10^{20}$ eV. The overall spectrum falls approximately as $E^{-3}$, as shown in Figure~\ref{fig:allspec}. The main spectral features are a steepening at few PeV, called the knee, and a softening at few EeV (1 EeV $=10^{18}$ eV), called the ankle. There is also a strong steepening close to $10^{20}$ eV. Up to $\sim$~TeV energies it is possible to detect directly the particles from space using satellites, balloons and detectors in the International Space Station, and precise determinations of the particle mass and charge are possible. At higher energies, the flux of particles is very low and CRs can only be detected indirectly through the air showers of secondary particles that they develop by interactions in the atmosphere.  
 
 The lower energy particles come from sources within our Galaxy, while at the highest energies they are of extragalactic origin. It is still not known at which energy happens the transition from Galactic to extragalactic CR origin.
 
 The main open questions that we would like to answer are where and how these cosmic particles have been accelerated and how they propagated to the Earth. We have to take into account that during their propagation the particles can interact with the radiation and matter present along their way and be deflected by the magnetic fields. Thus, the energy, arrival direction, and sometimes even the mass of the particles can be  changed.
 
 \section{Galactic Cosmic Rays}
 
 The standard scenario for Galactic CRs is that they are accelerated in shock waves in Supernova explosions and then propagate diffusively in the interstellar medium. Evidence for this comes from the diffuse $\gamma$-ray emission at GeV energies from the Galactic disk observed by Fermi LAT. This can be explained as the result of the decay of neutral pions produced in CR interactions with gas in the disk, $p+gas \rightarrow \pi^0 \rightarrow \gamma \gamma$. Moreover, several supernova remnants, such as IC433 and W44, show $\gamma$-ray emission from their surroundings with the characteristic spectral shape of the $\pi^0$ decay, that is evidence of hadrons being accelerated to CR energies in the supernova remnant shocks \cite{fermilat}.
 
 Galactic supernova remnants can provide enough energy to account for the observed CRs if $10\%$ of the kinetic energy of the explosion goes into accelerating CRs. Moreover, the diffusive shock acceleration mechanism predicts a spectrum for the accelerated particles close to a power-law, $\propto E^{-\alpha}$ with $\alpha\simeq 2$--2.4, that after taking into account the effects of a diffusive propagation in the Galactic magnetic field, is close to the observed one.
 
 The particles being accelerated are taken from the interstellar medium (ISM), so it is expected that CRs have a similar abundance of elements as the ISM. This is actually the case except for few nuclei: Li, Be, B, F and the Sc-Mn group, that are secondary particles produced by spallation of primary particles  colliding with matter in the Galactic disk. By studying the abundances of these secondary particles relative to the primary ones, as for example the B/C ratio, it is possible to study how was the propagation of the particles inside the Galaxy. It is found that particles spend a long time wandering around the Galaxy, propagating diffusively. Moreover, by studying the B/C ratio dependence with 
the rigidity of the particles ($\sim E/Z$ in the relativistic limit, with $Z$ the charge), it can be seen that the magnetic field turbulence is compatible with the Kolmogorov one \cite{amsbtoc}.
 
 We have seen in this conference many
 detailed measurements by the AMS Collaboration of the spectrum of individual elements (and even isotopes), $e^+$ and $e^{-}$, that gave an impressive input to the field of Galactic CRs. Explaining the observed spectral features led to a lively interplay between experiments and theories, from detailed modelling of the diffusion processes, investigating the need of new astrophysical sources and possible signals of dark matter \cite{wang}. These direct measurements of the  spectrum of individual nuclei are possible only up to 100~MeV - 1~TeV, depending on the nucleus. At higher energies, to study for example  the end of the Galactic CR spectrum and the extragalactic CRs, we have to resort to indirect measurements that cannot  make this individual measurements of the identity of the particles, and nuclei can only be grouped in broad ranges of masses.
 The results of KASCADE-Grande \cite{KGsp} show that the light component of the CRs has a steepening in its spectrum at energies close to the knee, while the heavy component has a steepening at around $10^{17}$~eV, where there is another feature of the spectrum that is called the second knee. These measurements can be interpreted as related to the end of the Galactic CR spectrum: the light component starts fading at the knee energy, while the heavy component starts fading at the second knee one, suggesting that the underlying cause for this steepening is related to the CR rigidities.
 
 \section{Indirect Cosmic Ray measurements}
 
 Cosmic rays of higher energy are detected through the air showers that they produce in the atmosphere. They can be measured using an array of detectors that sample the particles reaching the ground. The largest observatory in operation is the Pierre Auger Observatory located in the Province of Mendoza (Argentina). The array is composed of 1660 water-Cherenkov detectors in a 1.5~km triangular grid, covering $3000$~km$^2$. It is overlooked by 27 fluorescence detectors in four sites around the perimeter of the array. The largest observatory in the Northern hemisphere is the Telescope Array in Utah, USA, composed of 507 scintillators in a 1.2~km square grid, covering 700 km$^2$ and overlooked by 36 fluorescence detectors. The fluorescence detectors measure  the UV fluorescence light from nitrogen molecules excited by particles in the extensive air shower, leading to a glow along their path in the atmosphere.  Hybrid detectors work in the following way: in moonless nights, when the fluorescence detectors can be operated (with a duty cycle of $\sim 15\%$) the amount of deposited energy is measured as a function of the atmospheric depth, and by integrating it a calorimetric determination of the total energy deposited by the shower is obtained, from which the energy of the original particle can be estimated. The telescopes also measure the atmospheric depth at which the shower reaches its maximum development, which is a useful indicator of the mass composition of the incident particle.
 On the other hand, the array of surface detectors sample with nearly 100$\%$ duty cycle the particles reaching the ground at different distances from the shower core.
 This allows one to reconstruct the CR arrival direction and to estimate the signal expected at a reference distance, that for the Pierre Auger Observatory is chosen as 1000~m from the shower core. 
 This provides a good estimator for the energy of the incident particle, which can be calibrated with that measured by the fluorescence detectors for the fraction of the events measured by both detectors.
 
 \section{Spectrum and composition of ultra-high energy cosmic rays}
\begin{figure}[h]
\centering{
\includegraphics[scale=.25,angle=0]{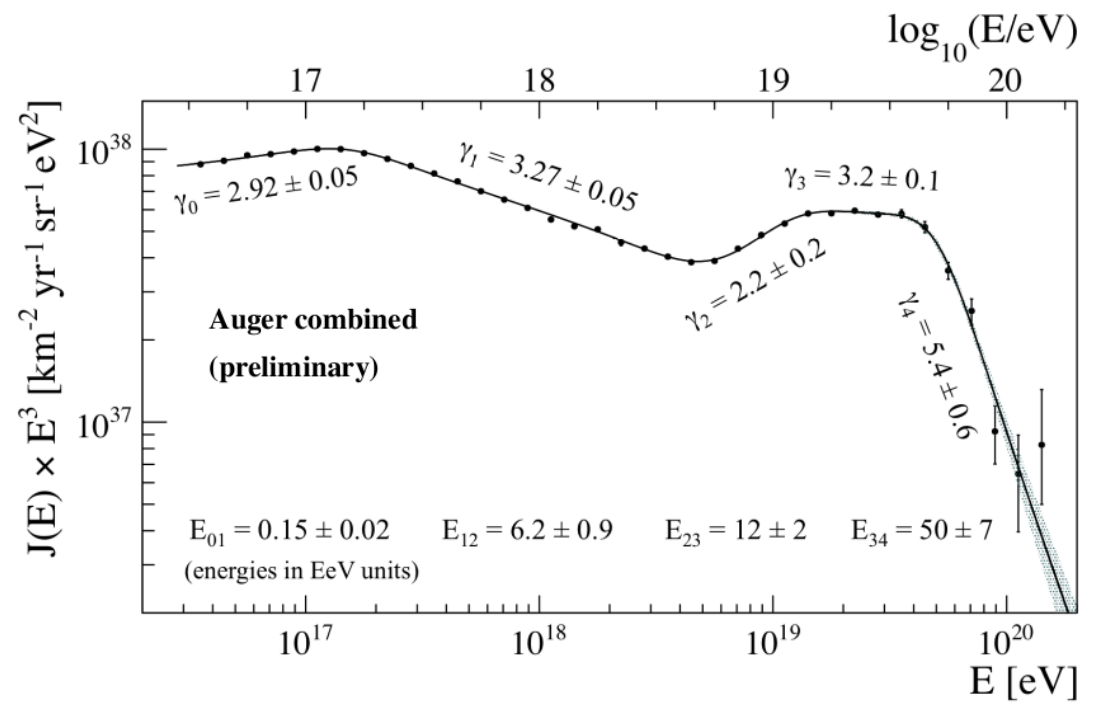}}
\caption{Spectrum of cosmic rays measured by the Pierre Auger Observatory, combining the results from the different detectors and analyses \cite{spicrc19}}
\label{fig:augersp}
\end{figure}

 The measured spectrum by the Pierre auger Observatory is shown in Figure~\ref{fig:augersp}. The second knee at 0.15~EeV, presumably due to the fading of the heavy mass galactic component, is clearly seen. At higher energies, the ankle hardening appears at 6.2~EeV, followed by a recently measured steeping at 12~EeV, before the very strong steepening observed at 50~EeV.
 
\begin{figure}[h]
\centering{
\includegraphics[scale=.4,angle=0]{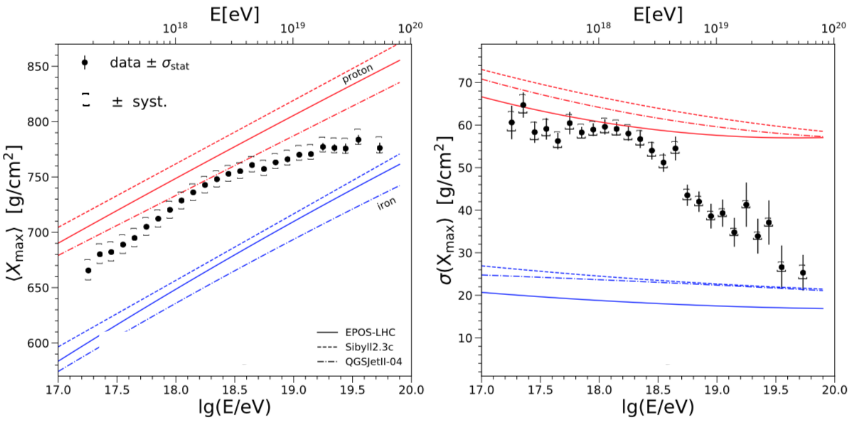}}
\caption{Measurements of the mean atmospheric depth of the shower maximum $X_{\rm max}$ and its dispersion from the Pierre auger Observatory \cite{comp}. The expectations for p and Fe primaries for different hadronic interactions models are also displayed.}
\label{fig:augercomp}
\end{figure}

The most widely used mass composition indicator for extensive air showers is the amount of atmosphere traversed up to the shower maximum, $X_{\rm max}$. Showers initiated by heavy nuclei develop higher in the atmosphere, and have smaller fluctuations than proton initiated ones. This can be understood by the superposition of smaller showers initiated by the nucleons. Figure~\ref{fig:augercomp} shows the expectations for p and Fe primaries for different hadronic interaction models together with the measurements by the Pierre Auger Observatory for the mean value of $X_{\rm max}$  (left panel) and its dispersion (right panel).
The data indicate that the composition becomes lighter for increasing energies up to 2~EeV, and then becomes heavier above this energy.
 
When we want to make contact between the spectrum and the composition of cosmic rays measured at the Earth and those of the accelerated particles at the sources, we have to take into account that CRs interact with the radiation backgrounds in their way, what makes them lose energy and even change composition. The main relevant processes  are pair production, photodisintegration of nuclei and photo-pion production. For example, at $10^{20}$~eV a proton or an Fe nucleus needs to come from distances closer than $\sim 75$~Mpc, and  nuclei with intermediate masses from even closer distances. 

\begin{figure}[h]
\centering{
\includegraphics[scale=.4,angle=0]{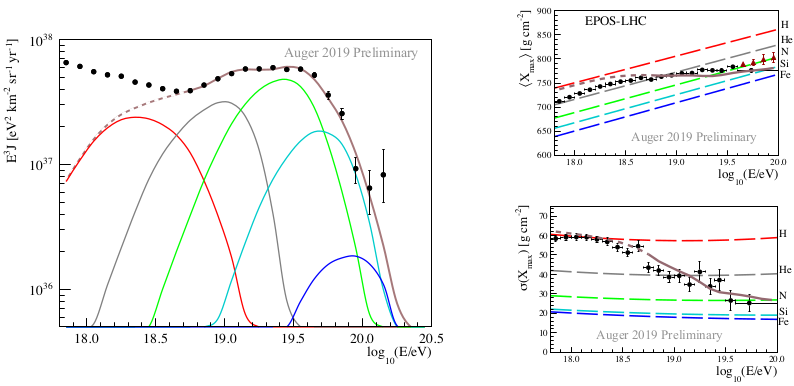}}
\caption{Fit to the cosmic ray spectrum and composition measurements above the ankle (solid brown lines) by the Pierre Auger Collaboration. In the left panel the contribution of the different mass groups (with the color code of the right panels) are displayed \cite{icrccast}}
\label{fig:augercombfit}
\end{figure}

If we try to interpret the observed spectrum and composition with a simple model of uniformly distributed  sources in which different elements are accelerated with a spectrum dependent on the rigidity of the particles, the Pierre Auger  Observatory results favour a scenario with mixed composition at the sources, with a very hard spectrum and a rather low rigidity cutoff $R_{cut}=E_{cut}/Z \simeq 5$~EeV. As shown in Figure~\ref{fig:augercombfit}, increasingly heavier elements dominate the spectrum with increasing energies \cite{combfitpap,icrccast}. In this kind of scenario,
the final steepening of the spectrum results from a combination of propagation effects and the maximum rigidity of the acceleration at the sources.
 
Let us notice that in the scenario described above to explain the spectrum and composition at the highest energies, this high energy component cannot give the dominant contribution to the spectrum at energies slightly below the ankle. Recalling that the Galactic component was fading above $10^{17}$~eV, there is a gap between the two contributions, so that there is probably the need for a new extragalactic component in the middle. 
Thus, it appears that two different extragalactic source populations would be required to be able to explain the full CR spectrum.

\section{Hadronic interactions in air showers}
We have already mentioned the electromagnetic part of the air showers, that is manly produced by the decays of neutral pions and that is the one measured by the fluorescence detectors,
and also contributes to the signals measured in the surface detectors. There is also a muonic part of the air showers, that comes mainly from the decays of charged pions.
The muonic component contains a subdominant fraction of the shower energy, but it can  give useful information about the primary particle mass composition \cite{schmidt}. Higher mass primaries produce showers with more muons than lighter ones. A good theoretical model for the development of air showers should be able to describe consistently both the electromagnetic and the muonic parts of the showers. Muons are measured at the Pierre Auger Observatory by an array of underground scinitillator detectors covering a small part of the surface array, that work in the energy range around $10^{18}$~eV. They can also be studied with the water-Cherenkov detectors by looking at the very inclined showers in which the electromagnetic part has been absorbed in the atmosphere, best suited in the energy range around $10^{19}$~eV. 

\begin{figure}[h]
\centering{
\includegraphics[scale=.38,angle=0]{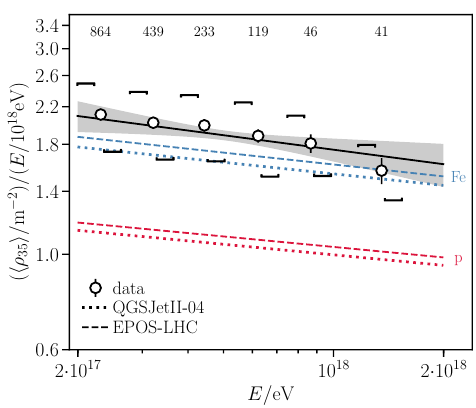}\includegraphics[scale=.28,angle=0]{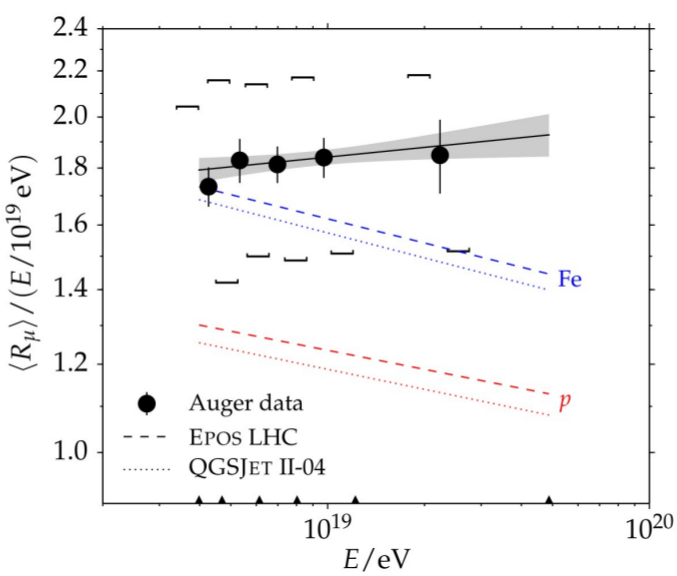}
\includegraphics[scale=.3,angle=0]{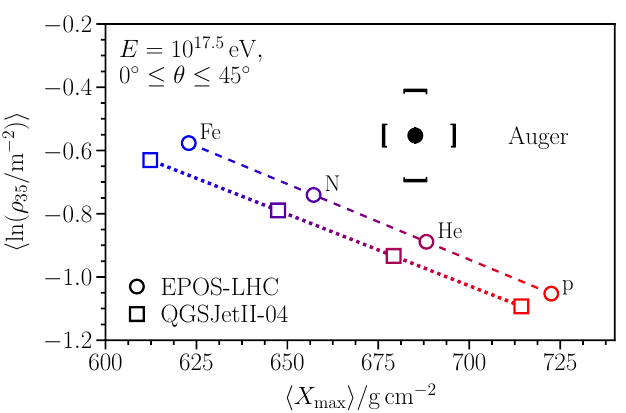}
\includegraphics[scale=.3,angle=0]{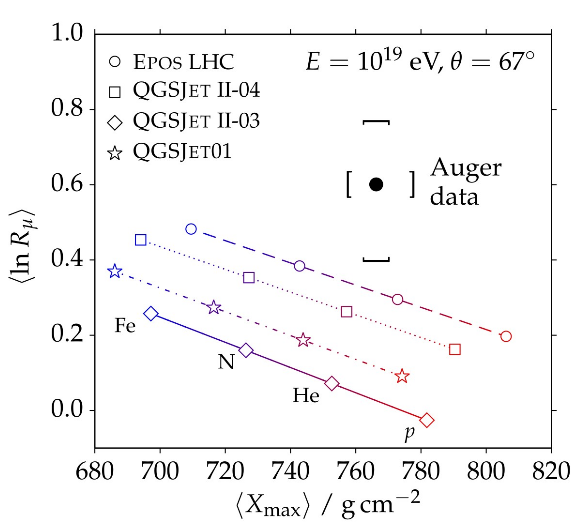}}
\caption{The upper panels show the muon density estimators for the underground muon detectors (left) \cite{umd} and for the inclined showers from the water-Cherenkov measurements as a function of the energy (right) \cite{hadintinc}. Lines corresponds to expectations for two different hadronic interaction models and points are the data. The lower panels show, for two different energies, the expected muon density estimator and mean $X_{max}$ for different composition and hadronic models (lines) and the measurements from the Pierre Auger Observatory.}
\label{fig:muons}
\end{figure}
 
The upper panels of Figure~\ref{fig:muons} show the expectations for the muon number estimators in the underground muon detectors (left) and for the inclined showers (right) \cite{hadintinc}, for proton and Fe showers and for different hadronic interaction models. Also shown are the data points, that turn out to be slightly above the expectations even for Fe composition.
Moreover, when we combine the $X_{\rm max}$ measurement with the muon measurements for the same energy bin, as shown in the lower panels for two energies, it is clear that none of the hadronic interactions models can consistently explain the observations for any composition \cite{schmidt}. 

Monte Carlo simulations of extensive air showers,  considering the mass composition inferred from the $X_{\rm max}$ measurements, predict a muon density at ground smaller than the observed one. There is a 30--50~$\%$ lack of muons in the simulations. This is an indication that some modification in the hadronic interaction models would be required .

\section{Arrival directions distribution}

\begin{figure}[h]
\centering{
\includegraphics[scale=.4,angle=0]{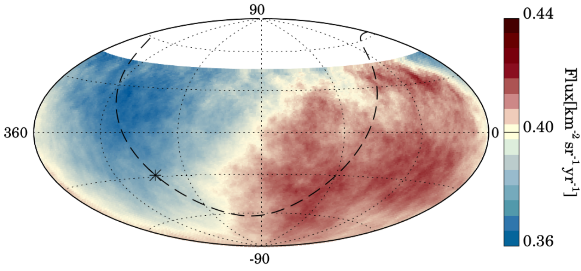}
\includegraphics[scale=.4,angle=0]{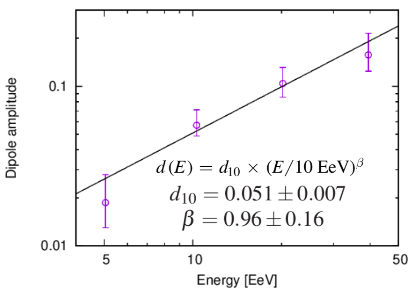}
}
\caption{Flux of CRs with energies above 8~EeV smoothed in an angular scale of $45^\circ$ (left panel) and amplitude of the dipolar component as a function of energy (right panel) \cite{dipicrc19}.}
\label{fig:dipole}
\end{figure}

The Pierre Auger Observatory has measured a dipolar anisotropy with an amplitude of $d=0.066^{+0.012}_{-0.009}$ in the arrival direction of CRs with energies above 8~EeV, with a significance of 6$\sigma$ \cite{dipicrc19}. The direction points to ($\alpha,\delta)=(98^\circ,-25^\circ)$, at $135^\circ$ from the direction of the Galactic center, this is evidence of an extragalactic origin. The left panel of Figure~\ref{fig:dipole} shows the flux smoothed in an angular window of $45^\circ$. The amplitude of the dipole is observed to be increasing with energy, as can be seen in the right panel of Figure~\ref{fig:dipole}. This is consistent with expectations, since higher energy particles travel more rectilinearly.
Telescope Array in the Northern hemisphere has recently measured a modulation in right ascension of amplitude $r=0.033\pm 0.019$, pointing to $\alpha=131^\circ \pm 33^\circ$ \cite{dipoleta}. This amplitude corresponds to an equatorial dipole component $d_\perp=r/\langle \cos\delta\rangle \simeq 1.3~ r\simeq 0.043\pm 0.025$, that is consistent both with the the Pierre Auger Observatory results ($d_\perp= 0.060\pm 0.010$) and with isotropic expectations within the statistical uncertainties. 
 
\begin{figure}[h]
\centering{
\includegraphics[scale=.25,angle=0]{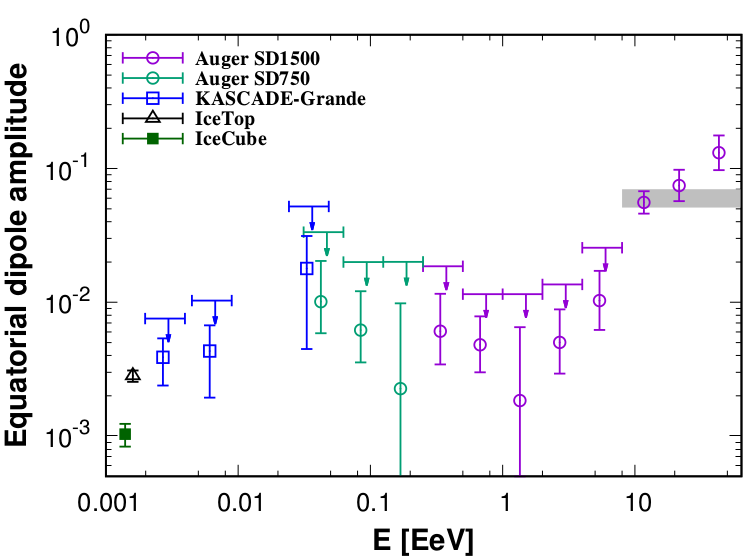}
\includegraphics[scale=.27,angle=0]{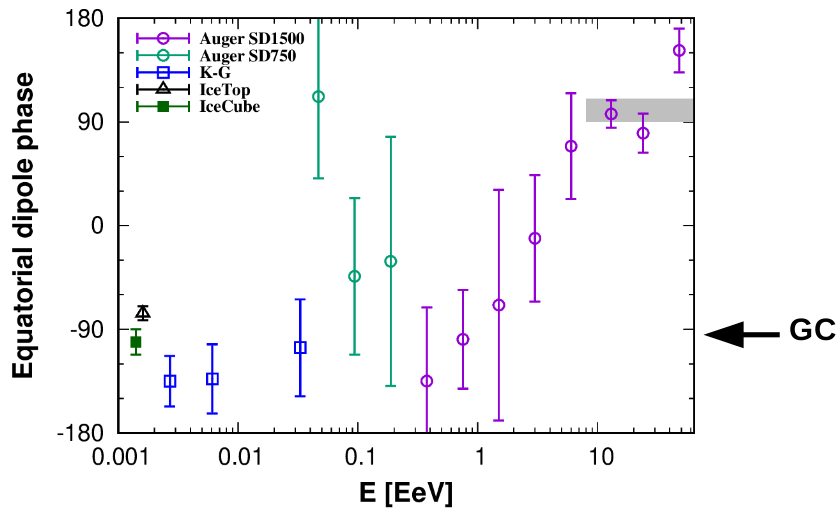}
}
\caption{Amplitude (left panel) and phase (right panel) of the equatorial component of the dipole from different experiments \cite{dipicrc19}. }
\label{fig:dperp}
\end{figure}

 Regarding the large scale anisotropies at lower energies, we show in Figure~\ref{fig:dperp} a compilation of the measurements by IceCube, KASCADE-Grande and the Pierre Auger Observatory of the equatorial dipole component in the energy range from 1~PeV to 100~EeV. The amplitude increases from few $10^{-3}$ to about 10$\%$ in that energy range, and the phases show a change from being close to the RA of the Galactic center at low energies to nearly the opposite direction at the highest energies, suggesting a transition from anisotropies of Galactic origin below $\sim 1$~EeV to extragalactic origin above few EeV.
 
\begin{figure}[h]
\centering{
\includegraphics[scale=.35,angle=0]{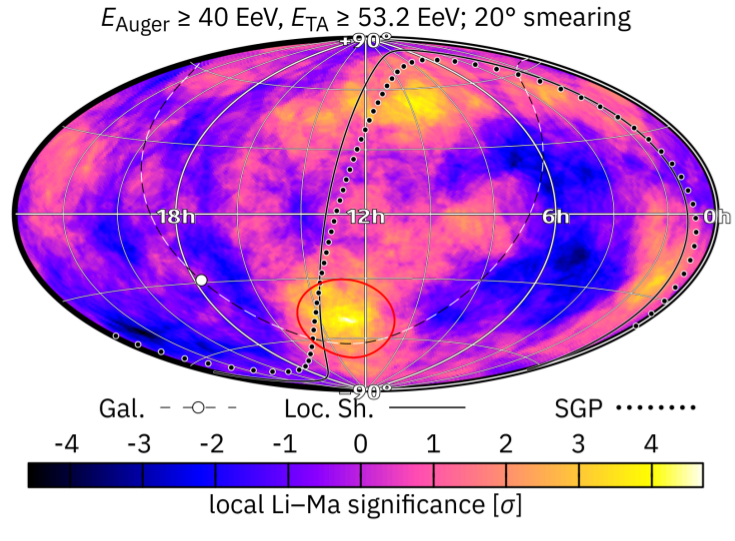}
}
\caption{Significance of the flux excess in a joint analysis of the Pierre Auger Observatory and Telescope Array data \cite{aniaugerta}.}
\label{fig:od}
\end{figure}

At higher energies and smaller angular scales, a combined analysis of the Pierre Auger Observatory and the Telescope Array data, allowing a full coverage of the sky, shows two hot spots, as seen in Figure~\ref{fig:od}. Different energy thresholds (40~EeV for Auger and 53.2~EeV for TA) were chosen to match the measured integral fluxes in the common observed region of the sky, and after scanning in 7 different angular window scales, the two most statistically significant excesses appeared at an angular window of $20^\circ$ around ($\alpha,\delta)=(192^\circ,-50^\circ)$, with 4.7~$\sigma$ pre-trial (2.2~$\sigma$ post-trial) and at an angular window of $15^\circ$ around ($\alpha,\delta)=(142^\circ,54^\circ)$, with 4.2~$\sigma$ pre-trial (1.5~$\sigma$ post-trial), where the post-trial significance accounts for the scan in angles and directions in the sky.

In summary, the arrival directions distribution of ultra-high energy cosmic rays shows a significant dipole above 8~EeV, there are some hints of medium scale anisotropies above 40~EeV and no evidence of small scale anisotropies. This is probably due to the fact that the deflections in the intergalactic magnetic field are significant, what is consistent with the transition to a heavier composition at the highest energies. This fact makes it difficult to  identify the actual sources. There are several proposed source candidates that meet the minimal conditions for being accelerator sites, like AGNs, starburst galaxies, GRBs and magnetars, but the identification is difficult since the arrival directions of CRs do not point to their sources due to the magnetic deflections.

\section{Multimessenger astronomy}

One possibility to try to identify the sites of CR acceleration  is to resort to other neutral messengers, like neutrinos, photons and neutrons (see e.g. \cite{zehrer}). Combined efforts in this direction are taking place, like the search for high energy neutrinos in coincidence with the binary neutron star merger GW170817 observed by Advanced LIGO and Virgo detectors in coincidence with a short GRB \cite{bns} (detected by Fermi-GBM and INTEGRAL and followed by many optical observatories around the globe). IceCube, Antares, the Pierre Auger Observatory \cite{gwmm} and also Baikal \cite{baikal} have looked for neutrinos in windows of $\pm 500$~sec and in the subsequent 14 days time and found none, putting upper limits to the flux of neutrinos from this source.

A very exciting recent result is the measurement of a high energy neutrino of 290~TeV detected by IceCube in coincidence with the direction of a rotating supermassive black hole, the blazar TXS 0506+056 at a redshift of 0.33, that was in an active flaring phase at that moment as measured by Fermi-LAT \cite{txs}. This is probably the first detection of an extragalactic cosmic ray accelerator.

\section{The Future}

To end this very quick review let us point out that at this moment there are improvements taking place in the largest CR observatories. The Pierre Auger Observatory is implementing an upgrade, AugerPrime, with the installation of plastic scintillators and radio antennas above all the water-Cherenkov detectors, with the aim to improve the sensitivity to mass composition 
by measuring the muonic and electromagnetic components of every shower \cite{cataldi}. The
Telescope Array is extending the area of the array  (TAx4) to cover 3000~km$^2$. In this conference we also heard about the plans to observe fluorescence light from space (POEMMA) \cite{poemma} and to instrument huge surfaces with radio antennas (GRAND) \cite{grand} that will boost the exposure at the highest energies in the next decade.

In summary, there have been many advances in understanding high energy cosmic particles.
There are still many open questions and new projects being developed and planned to answer them.

\end{document}